\documentstyle[sprocl,axodraw,graphicx,epsfig]{article}

\def\be{\begin{equation}}
\def\ee{\end{equation}}
\def\bea{\begin{eqnarray}}
\def\eea{\end{eqnarray}}

\begin{document}

\title{NON-LEPTONIC TWO BODY B DECAYS IN QCD}

\author{Cai-Dian L\"u}

\address{Institute of High Energy Physics,
P.O. Box 918(4), Beijing 100039, China  \\E-mail: lucd@ihep.ac.cn}


\maketitle\abstracts{ We review the current status of theoretical
study of non-leptonic two body B decays. There are two independent
directions for this purpose. One is the so called QCD
factorization approach (or BBNS approach), which is based on naive
factorization approach. The other one is named perturbative QCD
approach. We list the different ideas and applications of the two
method, and make a comparison between the two. }

\section{Introduction}

The current running B factories at KEK and SLAC arouse many
interests of theoretical studies of B decays. The mechanism of CP
violation and signal of new physics are among the most important
issues in B physics. Most of B topic involve the study of
non-leptonic B decays, such as CKM angle measurements and rare
decays etc. Quantum Chromodynamic is one of the successful theory
of particle physics, however, the non-perturbative behavior of QCD
is still the unsolved problem. Unlike perturbative QCD, the
hadronization of non-leptonic B decays can not be done from the
first principal. Thus it is model dependent. The most successful
approach recently is the factorization approach (FA), which can explain
many of the decay channels by very few parameters
\cite{fac,akl1,cheng}.

Although the predictions of branching ratios agree well with
experiments in most cases, there are still some theoretical points
unclear in FA. First, it relies strongly  on the form factors, which
cannot be calculated by FA itself. Secondly, the generalized FA
shows that the non-factorizable contributions are important in a
group of channels \cite{akl1,cheng}. The reason of this large
non-factorizable contribution needs more theoretical studies.
Thirdly, the strong phase, which is important for the CP violation
prediction, is quite sensitive to
 the internal gluon momentum \cite{akl2}.
This gluon momentum is the sum of  momenta of two quarks, which go
into two different mesons. It is difficult to define exactly in
the FA approach. To improve the theoretical predictions of the
non-leptonic $B$ decays, we try to improve the factorization
approach, and explain the size of the non-factorizable
contributions in a new way.

There are mainly two directions toward this improvement: One is
the so called QCD factorization or BBNS \cite{bbns}, which is
based on the naive factorization approach. In this approach, the non-factorizable
contribution can be directly calculated if it is not dominant over the factorizable one.
 The other is the perturbative
QCD (PQCD) approach which is based on Brodsky and Lepage's idea
\cite{bl}. In this direction, people try to calculate the form factors and also
the non-factorizable and annihilation type contributions in a systematic way.
In the next section, we will first explain the idea of
naive and generalized factorization approach, which has been
studied for years. In section 3, we will introduce the idea of QCD
factorization. And in section 4, PQCD approach is applied.
Finally, in section 5, we will give a brief comparison between the
two approaches and summarize at the end.

\section{Generalized Factorization Approach}

The calculation of non-leptonic decays involves the short-distance
Lagrangian and the calculation of hadronic matrix elements which
are model dependent.
 The short-distance QCD corrected Lagrangian is  calculated
 to { next-to-leading order}.
The popular method to calculate the  hadronic matrix elements is
using the factorization method where the matrix element is
expressed as a product of two factors
 $\langle h_1h_2 | {\cal H}_{eff}|B\rangle =\langle h_1 | J_1|B \rangle
\langle h_2 | J_2|0 \rangle $.    The first factor is proportional
to the $B\to h_1$ form factor, while the second one is
proportional to the decay constant of $h_2$ meson.

  First, let us discuss the short distance part. The non-leptonic
  decays of B mesons are induced by the weak interaction in the quark level, which
  gives the effective four-quark operator. Together with the QCD
  corrections,
the effective Hamiltonian for the charmless  non-leptonic B decays
is
\begin{eqnarray}
\label{heff} {\cal H}_{eff} = \frac{G_{F}}{\sqrt{2}} \, \left[
V_{q'b} V_{q'q}^* \, \left(\sum_{i=1}^{10} C_{i} \, O_i + C_g O_g
\right) \right]  ,
\end{eqnarray}
where $q=d,s$ and $V_{q'q}$ denotes the CKM factors.
The operators $O_1,O_2$ are tree level current operators. The operators
$O_3,\ldots,O_6$ are QCD penguin operators. $O_7,\ldots,O_{10}$
arise from electroweak penguin diagrams, which are suppressed by
$\alpha/\alpha_s$. Only $ O_9$ has a sizable value whose
 major contribution arises from the $Z$ penguin.

For example, let us consider B meson decays to two
 pseudoscalar mesons.
In the factorization approach, using the effective Hamiltonian,
 we  write the required matrix element in its factorized form
{\begin{eqnarray}
  \label{7}
&&\langle P_1 P_2 | {\cal H}_{eff} |  B \rangle 
=  i \frac{G_F}{\sqrt{2}} V_{qb}V_{q q'}^*{ a_i}
f_{P_2} (m_B^2-m_1^2) F_0^{B\to P_1} (m_2^2) .
\end{eqnarray}}
The dynamical details are coded in the  quantities $a_i$, which we
define as $a_i \equiv C_i^{eff} + C_j^{eff}/N_c$, where $\{i,j\}$
is any of the pairs $\{1,2\}$, $\{3,4\}$, $\{5,6\}$, $\{7,8\}$ or
$\{9,10\}$. In practice, $N_c$ is treated as a phenomenological
parameter to include the non-factorized color-octet contributions.

If the amplitude is
  dominated by the tree amplitude, the
BSW-classification  can be applied\cite{fac}.
 Class-I decays are color favored, whose matrix elements are
proportional to $a_1=C_1/N_c+C_2$.
 Class-II decays are color suppressed, whose
matrix elements are proportional to
 $a_2=C_1+C_2/N_c$. If $N_c=3$ as in QCD, then $a_2$ becomes a very small
 number, making the branching ratios of this class of decays also small. However
 from experiments we know that  in this category of decays, the branching ratios are
 not so small. Therefore the non-factorizable
 contributions are found to be very important in order to explain
 the large branching ratios.
  Class-III decays are  proportional to
  $a_1 + r a_2$. This class of decays will determine the relative
  sign between $a_1$ and $a_2$.
  The QCD  and electroweak penguins are also present in the
  charmless decays of B mesons.
For the penguin-dominant decays, we introduce two more classes \cite{akl1}:
Class-IV decays   involve one or more of the dominant penguin
coefficients $a_4$, $a_6$ and $a_{9}$.
 Class-V decays are decays with strong $N_c$-dependent
coefficients $a_3, a_5,a_7$ and $a_{10}$.

Class-I and Class-IV decays have relatively large branching ratios
of the order of $10^{-5}$ and stable against variation of $N_c$.
Most of the measured decay channels by experiments are belong to
these classes. There is a good agreement between experiments and
theory, which indicate the success of FA.
 Class-III decays are  mostly stable, except for some
 $B \to PV$ decays.
Class-II and Class-V decays are rather unstable against variation
of $N_c$. Some measured decay channels of these classes indicate
that we need an effective $N_c\simeq 2$ rather than $N_c = 3$ to
explain the experimental data. This implies that large non-factorizable
contribution exist in these decays.
 Many of them  may
receive significant contribution from the annihilation diagrams
and/or soft final state interactions.

In FA, the strong phase comes only from the so called BSS
mechanism \cite{bss}. The inner charm quark loop produce a strong
phase, when the charm quark on mass shell. This strong phase
depends on the inner gluon line strongly \cite{akl2}. Therefore,
in FA, the predicted CP violation  of B decays are in a quite
large range.

In the study of factorization approach, one finds that the
non-factorizable contributions are important in some of the
non-leptonic B decays, which can not be explained in the
factorization approach itself. Therefore theoretical study of this
effect is required.

\section{QCD Factorization Approach}

Recently, Beneke, Buchalla, Neubert, and Sachrajda (BBNS) proposed
a formalism for two-body charmless $B$ meson decays \cite{bbns}.
In this approach, they expand the hadronic matrix element by the
heavy b quark mass
\begin{equation}
\langle \pi \pi | Q | B \rangle = \langle \pi |j_1 |B\rangle
\langle \pi |j_2| 0\rangle \left[ 1+ \sum r_n \alpha_s^n + {\cal
O} (\Lambda_{QCD} /m_b) \right],\label{bbnsf}
 \end{equation}
 where Q is a local
operator in the effective Hamiltonian and $j_{1,2}$ are bilinear
quark currents. By neglecting the power corrections in
$\Lambda_{QCD}$, one need only calculate the order $\alpha_s$
corrections including the vertex corrections for the four quark
operators and the non-factorizable diagrams. These diagrams are
shown in Fig.1. The first 6 diagrams have already been included in
the generalized FA approach as next-to-leading order QCD
corrections to local four quark operators. What new are the last
two non-factorizable diagrams, which has a hard gluon line
connecting the four quark operator and the spectator quark.

They claimed that factorizable contributions, for example, the
form factor $F^{B\pi}$ in the $B\to\pi\pi$ decays, are not
calculable in PQCD, but nonfactorizable contributions are in the
heavy quark limit. Hence, the former are treated in the same way
as FA, and expressed as products of Wilson coefficients and
$F^{B\pi}$. The latter, calculated as in the PQCD approach, are
written as the convolutions of hard amplitudes with three
$(B,\pi,\pi)$ meson wave functions. Annihilation diagrams are
neglected as in FA. Hence, this formalism can be regarded as a
mixture of the FA and PQCD approaches. Values of form factors at
maximal recoil $q^2=m_\pi^2$ and nonperturbative meson wave
functions are all treated as input parameters. It is easy to see
from eq.(\ref{bbnsf}), that this equation is only applicable for
those color enhanced decay modes, where the factorizable
contribution dominates the final results. While for the color
suppressed modes, where the non-factorizable contributions   are
not small, the expansion of eq.(\ref{bbnsf}) is not right, since
the large non-factorizable contribution is grouped into the second
term of eq.(\ref{bbnsf}).

When extending the BBNS formalism to the $B\to D^{(*)}\pi$ decays,
 difficulty occurs in the calculation of nonfactorizable
amplitudes. The decay channel of $B^0  \to \pi^+ D^-$   is
calculable, because the end-point singularities in the
non-factorizable diagrams (last diagrams in Fig.1) cancel each
other. However, such a soft cancellation does not occur in the
calculation of $B^+ \to \pi^+ \bar D^0$ and $B^0\to \pi^0 \bar
D^0$ decays. The $D$ meson is heavy on the upper side of the
diagram, in which the light quark and the heavy $c$ quark do not
move collinearly. Therefore, soft gluons can resolve their color
structure, and interact with them.
 That is, not all the nonfactorizable
amplitudes in the BBNS formalism are calculable.

\begin{figure}[t]
\epsfig{file=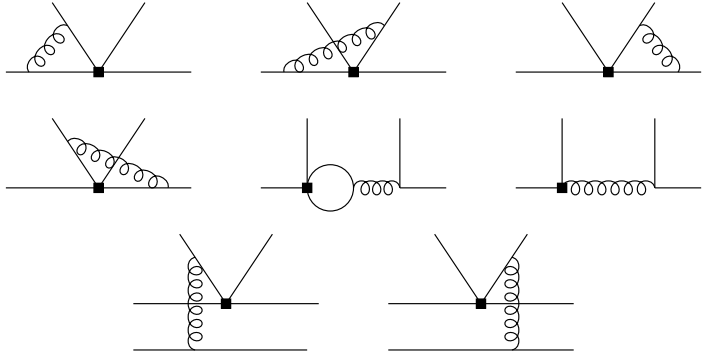,bbllx=2cm,bblly=19cm,bburx=11.5cm,bbury=25cm,%
width=8.3cm,angle=0} \caption{Figures calculated in the QCD
factorization approach. \label{fig:radish}}
\end{figure}

There are many calculations for various decay channels in this
approach. Group of people calculate $B$ meson decays to two light
pseudoscalar mesons \cite{bpp}, $B$ meson decays to one
pseudoscalar and one vector meson \cite{bpv} and $B$ meson decays
to final states with $\eta$ or $\eta^\prime$ \cite{etap} in the
QCD factorization approach. The numerical results show that the
theory and experiments agree well for those class I and IV decays,
which are color enhanced and dominated by the factorizable
contribution. This also agrees with the FA result, since the
dominant part in eq.(\ref{bbnsf}) is the same as the FA. The
success of QCD factorization is that one can calculate the
sub-leading ${\cal O} (\alpha_s)$ non-factorizable contribution
(second term in eq.(\ref{bbnsf})) using perturbative QCD. While in
the FA, one has to input a free parameter $N_c^{eff}$ to
accommodate the non-factorizable contribution.

  Cheng and  Yang did
the calculation of $B$ meson decay to two vector mesons
\cite{bvv}, including $ B\to \phi K^*$. In the calculation of
$B\to \phi K$ decay, they found that the annihilation
contributions are important, but the relative strong phase can not
be predicted exactly, due to the sensitivity of the cut-off
introduced \cite{bphik}. In fact, they found that a real
annihilation contribution is required to enhance the $B\to \phi K$
decay branching ratio. In the calculation of $B\to J/\psi K$
decay, they found that the leading twist contribution is too small
to accommodate the experimental data \cite{bpsik}. Problem remains
for color suppressed decay modes in QCD factorization approach.

In the calculations above, those people found that there exist
endpoint divergence in the annihilation diagram calculations of
QCD factorization \cite{du}. Logarithm divergence occurred at twist 2
contribution, and linear divergence exists in twist 3
contribution. If not symmetric wave function, like $K^{(*)}$ meson,
there is also soft divergence in the non-factorizable diagrams. It
is very difficult to treat these singularity in the BBNS approach.
A cut-off is introduced to regulate the divergence, thus makes the
QCD factorization approach prediction parameter dependent,
especially for the strong phase. Recently an effort is made to
introduce the $k_T$ dependence of the wave functions, and Sudakov
form factors in the BBNS approach in order to remove the singularities
 \cite{suda}. This makes the BBNS
approach to go toward the  direction of PQCD approach.

As for the strong phase in BBNS, like in FA, it may come from the
BSS mechanism. Here the momentum of the inner gluon is well
defined. However, it predicts too small strong phase, because of
the small gluon momentum. There is also another source of strong
phase from the annihilation diagrams, but strongly depends on the
cut-off parameter. The strong phase  in QCD factorization can be
almost arbitrary large.

Finally, the QCD factorization approach is at least one step
forward from Naive Factorization approach. It gives systematic
prediction of sub-leading non-factorizable contribution for the
class I and class IV decays, which are dominated by the
factorizable contribution.  Big problem is the endpoint
singularity, but may be solved with Sudakov from factors like PQCD
approach. The input parameters in QCD factorization approach are
form factors, wave functions etc.

\section{Perturbative QCD Approach}

In this section, we will introduce the idea of PQCD  approach \cite{litalk}.
The three scale PQCD factorization theorem has been developed for
non-leptonic heavy meson decays \cite{li}, based on the formalism
by Brodsky and Lepage \cite{bl}, and Botts and Sterman \cite{bs}.
In the non-leptonic two body B
 decays, the $B$ meson is heavy, sitting at rest.
It decays into two light mesons with large momenta. Therefore the
light mesons are moving very fast in the rest frame of $B$ meson.
In this case, the short distance hard process dominates the decay
amplitude. The reasons can be ordered as: first, because there are
not many resonance near the energy region of $B$ mass, so it is
reasonable to assume that final state interaction is not important
in two-body $B$ decays. Second, With the final light mesons moving
very fast, there must be a hard gluon to kick the light spectator
quark (almost at rest) in the B meson to form a fast moving light
 meson. So the dominant diagram in this theoretical picture
is that one hard gluon from the spectator quark connecting with
the other quarks in the four quark operator of the weak
interaction. Unlike the usual factorization approach, the hard
part of the PQCD approach consists of six quarks rather than four.
We thus call it six-quark operators or six-quark effective theory.
There are also soft (soft and collinear) gluon exchanges between
quarks. Summing over those leading soft contributions gives a
Sudakov form factor, which suppresses the soft contribution to be
dominant. Therefore, it makes the PQCD reliable in calculating the
non-leptonic decays. With the Sudakov resummation, we can include
the leading double logarithms for all loop diagrams, in
association with the soft contribution.


There are three different scales in the B meson non-leptonic
decay. The QCD corrections to the four quark operators are usually
summed by the renormalization group equation \cite{buras}. This
has already been done to the leading logarithm and next-to-leading
order for years. Since the $b$ quark decay scale $m_b$ is much
smaller than the electroweak scale $m_W$, the QCD corrections are
non-negligible. The third scale $1/b$ involved in the $B$ meson
exclusive decays is usually called the factorization scale, with
$b$ the conjugate variable of parton transverse momenta. The
dynamics below $1/b$
 scale is regarded as being completely
non-perturbative, and can be parameterized into meson wave
functions. The meson wave functions are not calculable in PQCD.
But they are universal, channel independent. We can determine them
from experiments, and it is constrained   by QCD sum rules and
Lattice QCD calculations. Above the scale $1/b$, the physics is
channel dependent. We can use perturbation theory to calculate
channel by channel.

Besides the hard gluon exchange with the spectator quark, the soft
gluon exchanges between quark lines  give out the double
logarithms $\ln^2(Pb)$ from the overlap of collinear and soft
divergence, $P$ being the dominant light-cone component of a meson
momentum. The resummation of these double logarithms leads to a
Sudakov form factor $\exp[-s(P,b)]$, which suppresses the long
distance contributions in the large $b$ region, and vanishes as
$b> 1/\Lambda_{QCD}$. This form factor  is given to sum the
leading order soft gluon exchanges between the hard part and the
wave functions of mesons. So this term includes the double
infrared logarithms.  It is  shown in ref.\cite{luy} that $e^{-s}$
falls off quickly in the large $b$, or long-distance, region,
giving so-called Sudakov suppression. This Sudakov factor
practically makes PQCD approach applicable. For the detailed
derivation of the Sudakov form factors, see ref.\cite{7,8}.

With all the large logarithms resummed, the remaining finite
contributions are absorbed into a perturbative b quark decay
subamplitude $H(t)$. Therefore the three scale factorization
formula is given by the typical expression,
\begin{equation}
C(t) \times H(t) \times \Phi (x) \times \exp\left[ -s(P,b) -2 \int
_{1/b}^t \frac{ d \bar\mu}{\bar \mu} \gamma_q (\alpha_s (\bar
\mu)) \right], \label{eq:factorization_formula}
\end{equation}
where $C(t)$ are the corresponding Wilson coefficients, $\Phi (x)$
are the  meson wave functions and the variable $t$ denotes the
largest mass scale of hard process $H$, that is, six-quark
effective theory.
 The quark anomalous dimension $\gamma_q=-\alpha_s /\pi$ describes
the evolution from scale $t$ to $1/b$. Since logarithm corrections
have been summed by renormalization group equations, the above
factorization formula does not depend on the renormalization scale
$\mu$ explicitly.

        \begin{figure}[tbp]
\epsfig{file=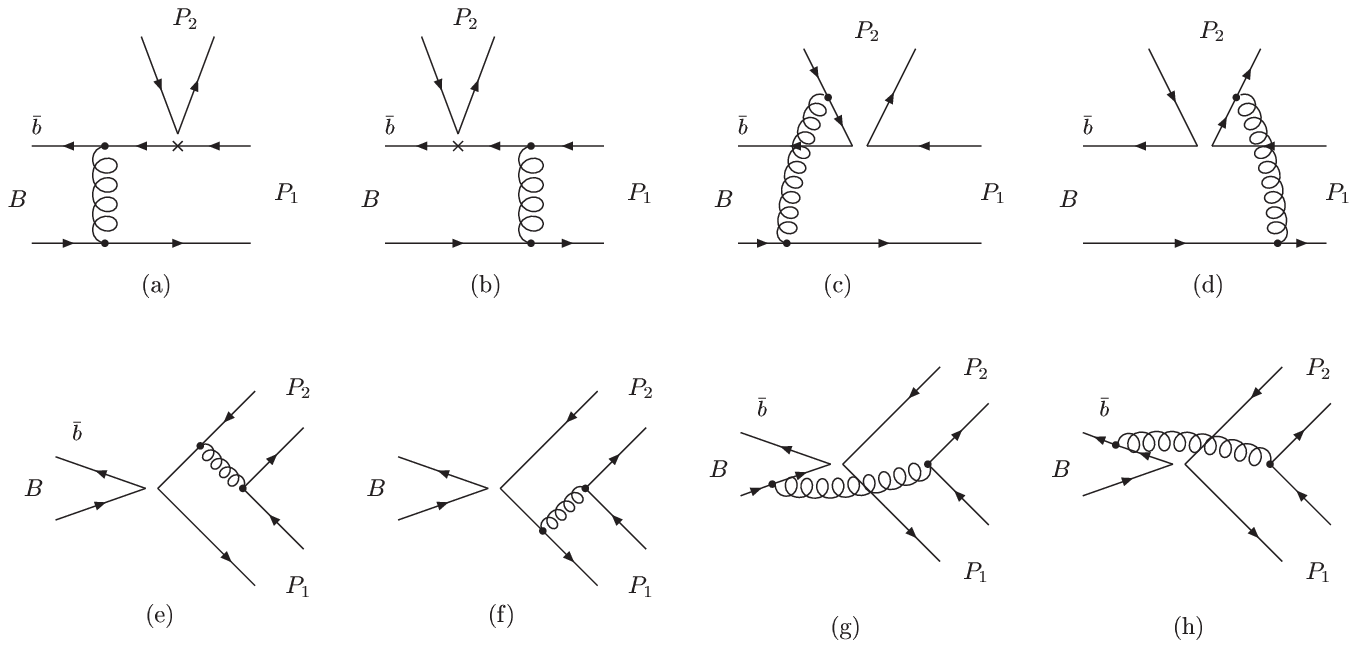,bbllx=4.6cm,bblly=18.9cm,bburx=15.5cm,bbury=25cm,%
width=9.6cm,angle=0}
    \caption{Diagrams for $B\to P_1P_2$ decay in perturbative QCD approach. The factorizable
    diagrams (a),(b), non-factorizable (c),
    (d),  factorizable annihilation
    diagrams (e),(f) and non-factorizable annihilation diagrams (g),(h).}
    \label{fig2}
   \end{figure}

The $\pi$ meson is treated as a light-light system. At the $B$
meson rest frame, pion is moving very fast. We define the momentum
of the pion which contain the spectator light quark as $P_2=
(m_B/\sqrt{2}) (1,0, {\bf 0}_T)$.  The light spectator quark
moving with the pion (with momentum $P_2$), has a momentum
$(k_2^+, 0, {\bf k}_{2T})$.
 If we define the momentum fraction as $x_2=k_2^+ / P_2^+$, then
the wave function of pion can be written as
\begin{equation}
\Phi_\pi= \frac{1}{\sqrt{2N_c}} \gamma_5 \left[\not \! p_\pi
\phi_\pi (x_2,{\bf k}_{2T}) + m_0 \phi^p_\pi (x_2,{\bf k}_{2T})
+m_0 \phi_\pi^\sigma (x_2,{\bf k}_{2T})\right], \label{eq:wf_pion}
\end{equation}
where $\phi_\pi (x_2,{\bf k}_{2T})$ is twist-2 wave function and
 $\phi^p_\pi (x_2,{\bf
k}_{2T})$ and $\phi_\pi^\sigma$ are twist-3 wave functions. The
$m_0$ in eq.(\ref{eq:wf_pion}) is given as
\begin{equation}
 m_0=\frac{m_\pi^2}{m_u+m_d}.
\label{eq:def_m0}
\end{equation}
It   is not the pion mass. Since this $m_0$ is a scale
characterizing the Chiral symmetry breaking,  it is estimated
around $1\sim 2$ GeV using the quark masses predicted from lattice
simulations, one may guess contributions of $m_0$ term cannot be
neglected because of $m_0 \not \!\ll m_B$. In the previous
calculation of $B\to \pi\pi$ \cite{luy} and $B\to \pi K$ decays
\cite{kls} we do not include the last term in
eq.(\ref{eq:wf_pion}) for the pion wave function. However, by
using a phenomenology twist 3 wave function for $\phi_\pi^p$, we
get the right result for those branching ratios. The reason is
that this choice of twist 3 wave function $\phi_\pi^p$,
accommodate the full twist 3 contribution effectively. This
phenomenological model also accommodate effectively the threshold
resummation effect discussed below.

 In the
PQCD approach, we can calculate not only the factorizable diagrams
(Fig.2(a),(b)) and non-factorizable diagrams (Fig.2(c),(d))
contribution but also the annihilation type diagrams
(Fig.2(e,f,g,h,)). Unlike the QCD factorization approach, there is
no logarithm or linear divergence in these calculations due to the
reason of Sudakov suppression with $k_T$ resummation and the
threshold resummation discussed above.

As shown above, in the PQCD approach, we keep the $k_T$ dependence
of the wave function. In fact, the approximation of neglecting the
transverse momentum can only be done at the non-endpoint region,
since $k_T \ll k^+$ is qualified  at that region. At the endpoint,
$k^+ \to 0$, $k_T$ is not small any longer, neglecting $k_T$ is a
very bad approximation. By, keeping the $k_T$ dependence,
 there is no endpoint divergence as
occurred in the QCD factorization approach, while the numerical
result does not change at other region. Furthermore, the Sudakov
form factors suppress the endpoint region of the wave functions.
Recently another type of resummation has been observed. The loop
correction to the weak decay vertex produces the double logarithms
$\alpha_s\ln ^2 x_2$ \cite{threshold}. Using the wave functions
from light-cone sum rules, at the endpoint region, these large
logarithms are important, they must be resummed. The threshold
resummation for the jet function results in Sudakov suppression,
which decreases the contribution of endpoint region of wave
functions. Therefore, the main contributions to the decay
amplitude in PQCD approach comes not from the endpoint region. The
perturbative QCD is applied safely.

 By including $k_T$ to regulate the
divergence, large logarithmic corrections $\alpha_s\ln k_T$
appear, and Sudakov resummation is demanded. With the resultant
Sudakov suppression, we have explicitly shown that almost 100\% of
the full contribution to the $B\to\pi$ transition form factor
arises from the region with the coupling constant $\alpha_s/\pi
<0.3$ \cite{kls}. It indicates that dynamics from hard gluon
exchanges indeed dominate in the PQCD calculation. In \cite{bbns}
Sudakov resummation is irrelevant, since all QCD dynamics has been
parameterized into models of form factors.

We emphasize that nonfactorizable and annihilation diagrams are
indeed subleading in the PQCD formalism as $M_B\to \infty$. This
can be easily observed from the hard functions in appendices of
ref.\cite{kls,luy}. When $M_B$ increases, the $B$ meson wave
function enhances contributions to factorizable diagrams. However,
annihilation amplitudes, being independent of B meson wave
function, are relatively insensitive to the variation of $M_B$.
Hence, factorizable contributions become dominant and annihilation
contributions are subleading in the $M_B\to\infty$ limit
\cite{bkphi}. Although the non-factorizable and annihilation
diagrams are subleading for the branching ratio in color enhanced
decays, they provide the main source of strong phase, by inner
quark or gluon on mass shell. The BSS mechanism strong phase is
negligible in the PQCD approach. In fact, the factorizable
annihilation diagrams are Chirally enhanced. They are not
negligible in PQCD approach \cite{kls,luy}. In the decays of B
meson to two light mesons, we collect terms up to chirally
enhanced terms ($O(m_0/m_B)$),  but still drop the terms
suppressed by $\Lambda/m_B$ \cite{bkphi}.

The main input parameters in PQCD are the meson wave functions.
It is not a surprise that the final results are   sensitive to the
meson wave functions. Fortunately, there are many channels involve
the same meson, and the meson wave functions should be process
independent. In all the calculations of PQCD approach, we follow
the rule, and we find that   they can explain most of the measured
branching ratios of B decays.   For example: $B\to \pi\pi$ decays
  \cite{luy}, $B\to K\pi$ decays \cite{kls}, $B\to \pi \rho$,
  $B\to \pi \omega$ decays \cite{ly}, $B\to K K$ decays
  \cite{bkk}, the form factor calculations of $B\to \pi$, $B\to \rho$
  \cite{semi}, $B\to K \eta^(\prime)$ decays \cite{eta},
   $B\to K\phi$ decays \cite{bkphi}    etc.

\section{Summary}

For a comparison, in the QCD factorization: Form factors are
 input parameters, which are claimed to be dominated  by soft contribution and not calculable.
The endpoint singularity is a crucial point in this approach, a
 cut off is needed to regulate the divergence.
 The QCD factorization follows FA, in which it has been assumed
 that factorizable contributions, being the dominant contribution.
 And all other contributions such as
 non-factorizable and annihilation diagrams,    being $\alpha_s$
 corrections, are sub-leading. In the limit of $\alpha_s \to 0$, it goes back to FA.
 Therefore
 the current BBNS approach can not be applied to the
 non-factorizable dominant process such as
$B^0 \to D^0 \pi^0$, and also those annihilation diagram dominant processes.

In the PQCD approach, the form factors are  calculable, which are
 dominant by short distance contribution. By including the $k_T$ dependence and
Sudakov suppression, there is no endpoint divergence.
In the PQCD formalism
non-factorizable  amplitudes are of the same order
as factorizable ones in powers of $1/M_B$, which are both
$O(1/(M_B\Lambda_{\rm QCD}))$.
 The smaller magnitude of nonfactorizable amplitudes in color
 enhanced decays are
due to the cancellation of the two non-factorizable diagrams. From
the viewpoint of power counting, they are of the same order. In
case of $B\to D \pi$ decays, the cancellation is absent. The power
counting changes, so that we   can also calculate the
non-factorizable contribution dominant processes.

For a conclusion, non-Leptonic B decays are important in
B physics. The theory of non-leptonic B decays is a challenging
work. There are still problems to be solved. The QCD factorization
and PQCD approach
   are going toward the same object, although there are still some differences.
They will be tested by experiments soon.

\section*{Acknowledgments}
We thank  Y.Y. Keum, E. Kou, T. Kurimoto, H.N. Li, T. Morozumi,
A.I. Sanda,  K. Ukai and M.Z. Yang for helpful discussions.

\end{document}